\documentstyle[aps,twocolumn,amssymb,epsfig]{revtex}
\begin{document}
\author{Wojciech S\L{}OMI\'NSKI %\thanks{email: wojteks@thp1.if.uj.edu.pl}
 and Jerzy SZWED}
\address{Institute of Computer Science, Jagellonian University,
Reymonta 4, 30-059 Krak\'ow, Poland}
\title{RESOLVED ELECTRON$^*$}
%\date{\today}
\maketitle
%-----    Abstract    ------------------------------------------------
\begin{abstract}
Advantages of introducing the electron structure function in electron induced
processes are demonstrated. At present energies
the same experiment gives  more precise description of the electron than 
photon structure.  The momentum scales entering the process are
better controlled. At very high momenta probabilistic (partonic) 
interpretation
can be preserved despite strong $\gamma$ -$Z$ interference.
The ``virtual photon'' structure  can also be reformulated 
in  terms of  more physical
(real) electron variables.
\end{abstract}
%\pacs{ xxxxx}

Theoretical framework which allows to calculate  the  photon structure
is known since long \cite{resph}. It 
 appears as perturbative QCD contribution 
(resolved photon),
in addition to the modeled Vector Meson Dominance (VDM) 
and pointlike (direct) 
terms.
To measure  this photonic structure,  experiments   
\cite{resphexp} use the electron (or positron) beam
as a source of photons.
Despite precise measurement
the data are difficult to extract. The problem is displayed   
in Fig. 1a. The tagged (upper) electron emits a probing photon whereas the untagged
(lower) goes nearly along the beam, emitting the target photon.
The large scale, $Q^2$ is determined by the tagged electron:

\begin{equation}
$$Q^2=-(k-k')^2=2EE_{\rm tag}(1-\cos \theta_{\rm tag}),
\label{QQ}
\end{equation}
where $E$ is the initial electron energy and $E_{\rm tag}$ and $\theta_{\rm tag}$
are the energy and polar angle of the measured electron. The anti-tagging 
condition (if present) requires the virtuality of the target photon to be 
less than $P^2$:
$$
-(p-p')^2 \equiv P_\gamma^2 \le P^2.
$$
This photon is clearly not a beam particle and
has the energy diffused 
according to the equivalent photon (Weizs\" acker-Williams \cite{WW})
spectrum $f^e_{\gamma}$. 
The measured cross-section for the production of a hadronic system $X$, expressed in terms of the 
photon structure functions $F_2^{\gamma}$ and 
$F_L^{\gamma}$ reads:
\begin{eqnarray}
\label{fot}
&&{{d^2\sigma _{ee\rightarrow eX}}\over {dzdQ^2}}
= 
{{2\pi {\alpha ^2_{\rm em}}}\over {x^2Q^4}} 
\nonumber 
\\
&&\times [(1+(1-y)^2)F^{\gamma}_2(x,Q^2,P^2)-y^2F_L^{\gamma}(x,Q^2,P^2)]
\nonumber 
\\ &&\times f_{\gamma}^e(z/x,P^2)dx
\end{eqnarray}
where 
$$y=1-(E_{\rm tag}/E)\cos ^2(\theta_{\rm tag}/2)$$
and
$x$ ($z$)  are fractions of parton  momentum with respect
to the photon (electron).

Three remarks are important for further considerations. 
First, 
the splitting of the process into a distribution of photons inside electron $f_{\gamma}^e$ 
and that
of partons inside the photon $F_2^{\gamma}$ is an approximation. 
The optimal form of the 
equivalent photon formula is still being discussed \cite{equiv}. Even if most of 
experimental groups  choose the same 
formula, one should keep in mind
 that the definition of the
photon structure 
function depends on this convention.
 
Second,
in order to fix $x$, one is forced to measure --- in addition to the tagged 
electron --- the hadronic momenta. In fact, 
$$
x= {Q^2 \over Q^2+P_\gamma^2 + W^2}
 \approx
 {Q^2 \over Q^2 + W^2},
$$
where 
$W$ is the invariant mass of the produced hadronic system $X$. Its determination 
is more difficult than other (tagged electron) variables.

Third, the target photon is 
always off-shell. 
Although the equivalent photon distribution is peaked at minimum (nearly zero)
virtuality, treating the photon as 
real  is another approximation. 
One should keep in mind that the measured photon structure function depends on 
$x$, $Q^2$ and $P^2$ 
(we keep the minimum virtuality fixed but in general it is still another variable). 

%%%%%%%  fig.1 **********************************
\begin{figure}
\vspace{2mm}
\centerline{\epsfig{file=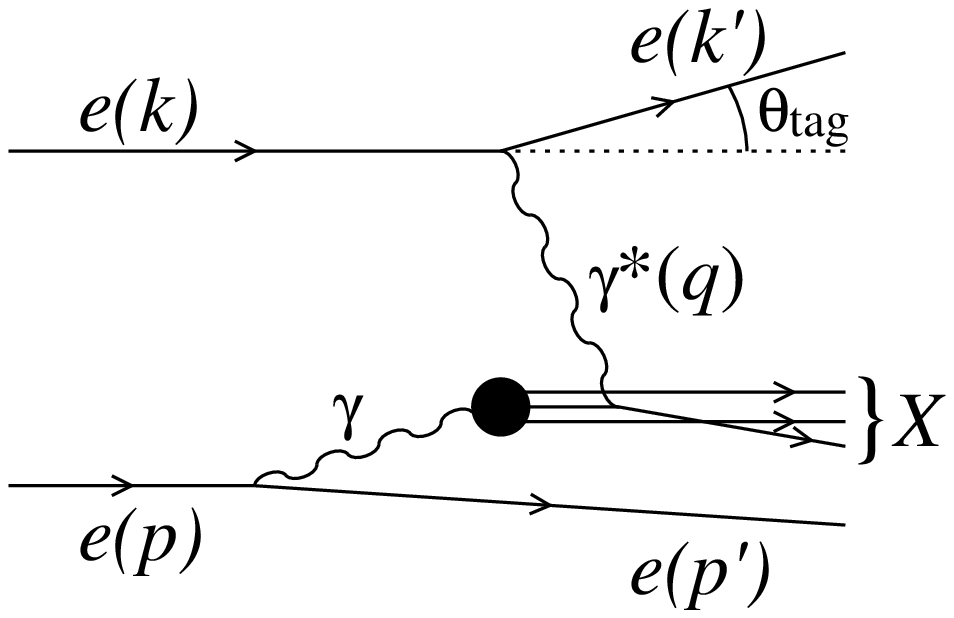,width=6.5cm}}
\centerline{\large (a)\hfil}
\vspace{5mm}
\centerline{\epsfig{file=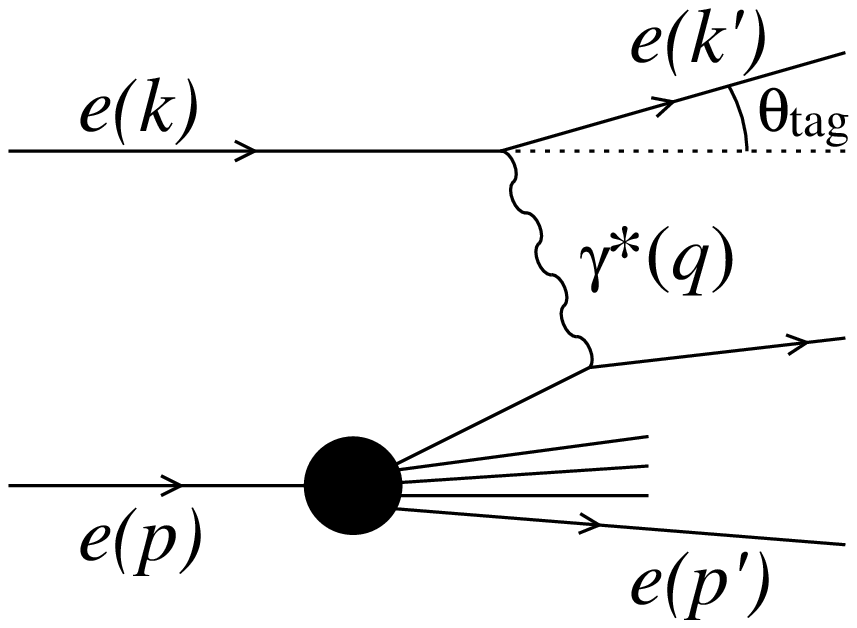,width=6.5cm}}
\centerline{\large (b)\hfil}
\vspace{5mm}
\caption{Deep inelastic scattering on a photon (a) and electron (b) target}
\end{figure}

The necessity of unfolding the photon structure function from the
cross-section (\ref{fot}) and
the uncertainty in  the determination of the
$x$ variable are sources of biggest errors in the analysis. 
The data are indirectly biased by theoretical assumptions and, in addition,
the selection cuts put on the hadronic mass $W^2$ reduce the number of accepted
events.

Most of the above problems can be avoided when we introduce the
structure function of the electron (Fig. 1b). To see how it works 
let us first write the cross-section (\ref{fot}), this
time in terms of the electron structure functions $F_2^e$ and $F_L^e$:
\begin{eqnarray}
\label{ele}
&&{{d^2\sigma _{ee\rightarrow eX}}\over {dzdQ^2}} =
 {{2\pi {\alpha ^2_{\rm em}}}\over
{z Q^4}}
\nonumber \\
&&[(1+(1-y)^2)F^e_2(z,Q^2,P^2)-y^2F_L^e(z,Q^2,P^2)].
\end{eqnarray}

The structure function $F_2^e(z,Q^2,P^2)$ which  dominates the
cross-section at  small $y$, has simple partonic 
interpretation:
$$
F_2^e(z,Q^2,P^2)=z\sum_i e^2_{q_i} q_i(z,Q^2,P^2)
$$ 
where $e_{q_i}$ and  $q_i$ are the $i$-th quark fractional charge and density.
 No unfolding
procedure is necessary to obtain $F_2^e$, and its argument $z$ --- the parton momentum fraction
with respect to the electron --- is measured, as in the standard deep inelastic 
scattering, with the tagged electron
variables only:
$$
z={Q^2\over {2pq}}=
%{1-\cos\theta_{\rm tag} \over 2E/E_{\rm tag} -1 -\cos\theta_{\rm tag}}.
{\sin^2(\theta_{\rm tag}/2) \over E/E_{\rm tag} -\cos^2(\theta_{\rm tag}/2) }.
$$
There is no need to reconstruct the hadronic mass $W$.
All these features cause that
 the same experiment can produce more precise data. 
What is most important
--- the electron structure function contains the same information about 
QCD  as the photon one
and, as we point out below, it is known theoretically with at least the same accuracy. 
Moreover, it allows to avoid  
problems which arise in the photon structure function
at very high energies. 

The construction of the QCD electron structure function can be presented 
in two steps.
First we calculate the splitting function of the electron into
a quark/anti-quark
$P^e_{q/\bar q}(z,P^2)$ (Fig. 2). 
As a high $Q^2$ probe we use in this calculation the gluon rather than the photon, 
because it couples to the QCD partons only.
Note that we 
allow for the exchange of $Z$ and $W$ bosons in addition to the photon.
The result for a quark reads 
(for anti-quark: 
exchange $\Phi_+$ with $\Phi_-$ and replace $\delta_{qd}$ with $\delta_{\bar q \bar u}$):
\begin{eqnarray}
\label{evol}
P^e_q(z,P^2)={{3\alpha_{\rm em}}\over{4\pi}}
\Big\{
2e_q^2 \left[\Phi_+(z)+\Phi_-(z)\right]  \log\mu_0 
&&
\\
+\tan^4\theta_{\rm W}\left[(e_q^2+z_q^2\rho^2)\Phi_+(z)+(e_q^2\rho^2+z_q^2)
\Phi_-(z)\right] && \log\mu_Z
\nonumber \\
+2 e_q \tan^2\theta_{\rm W} \left[(e_q+z_q\rho)\Phi_+(z)
-(e_q\rho+z_q)\Phi_-(z)\right] && \log\mu_Z
\nonumber \\
+(1+\rho)^2\Phi_+(z)\delta_{qd} && \log\mu_W
\Big\}
\nonumber
\end{eqnarray}
where
\begin{eqnarray}
&&\Phi_+(z)={{1-z}\over {3z}}(2+11z+2z^2)+2(1+z)\log z,\\
&&\Phi_-(z)={{2(1-z)^3}\over {3z}}
\end{eqnarray}
$\rho={1/{2\sin^2\theta_{\rm W}}}-1,\,
\mu_0=P^2 / m_e^2,\,
\mu_B=P^2/M_B^2 +1,\,
z_q={T_3^q / \sin^2\theta_{\rm W}}-e_q
$,
with $e_q$, $T_3^q$, $M_B$, $m_e$ and $\theta_{\rm W}$ being the quark charge, 
its 3$^{\rm rd}$ weak isospin 
component, weak boson ($Z$ or $W$) mass, electron mass and the Weinberg angle.
One sees that at present $Q^2$ the photon contribution (first term proportional to $\log \mu_0$) dominates. At very high momenta one should expect
the other terms to contribute, in particular the $\gamma$-$Z$ interference
enters with the same logarithm as the $Z$-term itself.

%%%%%%%  fig.2 **********************************
\begin{figure}
\centerline{\epsfig{file=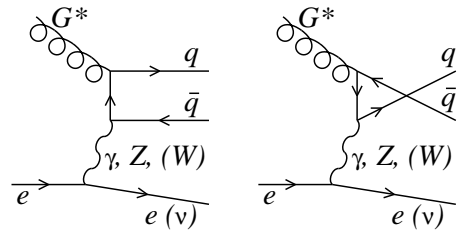,width=6cm}}
\vspace{2mm}
\caption{The Feynman diagrams contributing to the 
electron $\rightarrow$ quark/anti-quark splitting functions}
\end{figure}

In the second step we construct the $Q^2$-evolution equations for the quark and gluon 
densities inside the electron $q(z,Q^2,P^2)$ and $G(z,Q^2,P^2)$. 
Introducing
$t=\log (Q^2/\Lambda_{\rm QCD}^2)$ and $t_1=\log (P^2/\Lambda_{\rm QCD}^2)$ and 
remembering that there is no
direct coupling of the electroweak sector to gluons we can write \cite{WS22}:
\begin{eqnarray}
\label{master}
{{dq_j}(t,t_1)\over dt} &=& {\alpha_{\rm em}\over  2\pi} 
  P_{q_j}^e(t_1)
\nonumber\\
  &+& {\alpha(t)\over 2\pi} \sum_kP_{q_j}^{ q_k} \otimes q_k(t,t_1)
  +P_{q_j}^{G} \otimes G(t,t_1),
\\
\label{master1}
{dG(t,t_1) \over dt} &=&
 {\alpha(t)\over 2\pi} \sum_k P_{G}^{q_k} \otimes
q_k(t,t_1) + P_{G}^{G} \otimes
G(t,t_1).
\label{master2}
\end{eqnarray}
The sign $\otimes$ stands for the convolution 
$f\otimes g=\int_z^1 {dx\over x} f(z/x)g(x)$ and  explicit $z$  dependence has 
been suppressed.
$\alpha(t)$ is the QCD running coupling constant.

The momentum scales  require some attention.  The QCD evolution, started
by the $q\bar q$ pair, is governed by $\Lambda_{\rm QCD}$. The 
large `probing scale' defined for  $e^+e^-$ scattering in Eq.(\ref{QQ}),
is  chosen in  other reactions to be large transverse momentum (e.g. of  jets) 
or  
produced heavy mass (e.g. heavy quark or Higgs boson). The maximum virtuality 
$P^2$
defines the kinematical range which is called photo- (or more general
 boso-)
production. It
is often set by experiment. E.g. in   $e^+e^-$ or $ep$ scattering 
it can be chosen by the untagging condition.  But one may also take into 
account larger virtualities up to 
$Q^2\lesssim P^2$. 
In   the case of the $W$ boson contribution no such limit can be imposed experimentally at all 
and $P^2$  varies up to its kinematical limit:
$$
P^2_{\rm max}={{x+1-z}\over x}Q^2.
$$
In the asymptotic $t$ region ($t \rightarrow \infty$) we can take \cite{wojtus}
$$
t_1=(1-a)\hat t_1 +at
$$
where $\hat t_1$ is constant and $0\le a\le 1$.
Let us look closer at the two limiting cases $a=0$ and $a=1$. In the
first one the inhomogeneous
term in the evolution equations (\ref{master}) does not depend on $t$.   
The asymptotic solutions have the form:
\begin{eqnarray}
q(z,t) &&\simeq
\left( {\alpha_{\rm em} \over 2\pi} \right)^2 
q^{\rm as}(z) \,  t_1 t,
\nonumber
\\
G(z,t) &&\simeq 
\left( {\alpha_{\rm em} \over 2\pi} \right)^2 
G^{\rm as}(z) \,  t_1 t
\label{asdef}
\end{eqnarray}
with $q^{\rm as}(z)$ and $G^{\rm as}(z)$ being given by known, $t$-independent
integral equations \cite{WS22,wojtus}.

Their numerical solutions are shown in Fig. 3.  As expected 
in the asymptotic 
region all bosons contribute. What is interesting, 
the $\gamma$-$Z$ interference term
enters with strength comparable to the contribution of the $Z$ boson itself.  
This means that 
the notion of separate equivalent bosons breaks down at
very high momenta. The 
electron structure function  takes correctly 
into account interference effects,
preserving at the same time  probabilistic (partonic) interpretation. 

One should keep in mind that only in the asymptotic region all logarithms
entering the splitting function $P^e_q$ are equal. At lower energies
the photon contribution (proportional to $\log \mu_0$) dominates.

%%%%%%%  fig.3 **********************************
\begin{figure}
\centerline{\epsfig{file=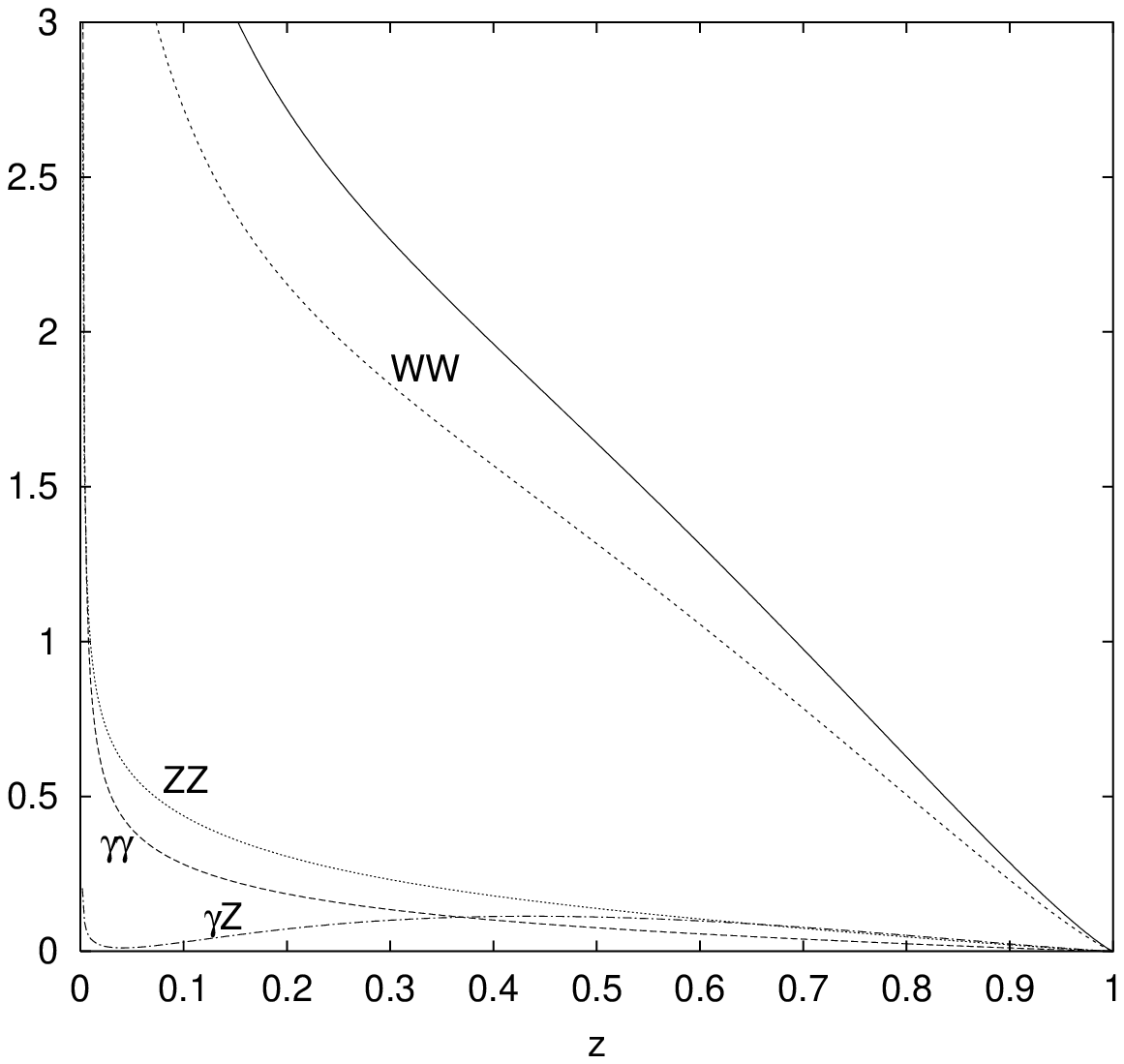,width=\columnwidth}%
\unitlength=1mm
\put(-25,65){$z d^{\rm as}(z)$}}
\centerline{\epsfig{file=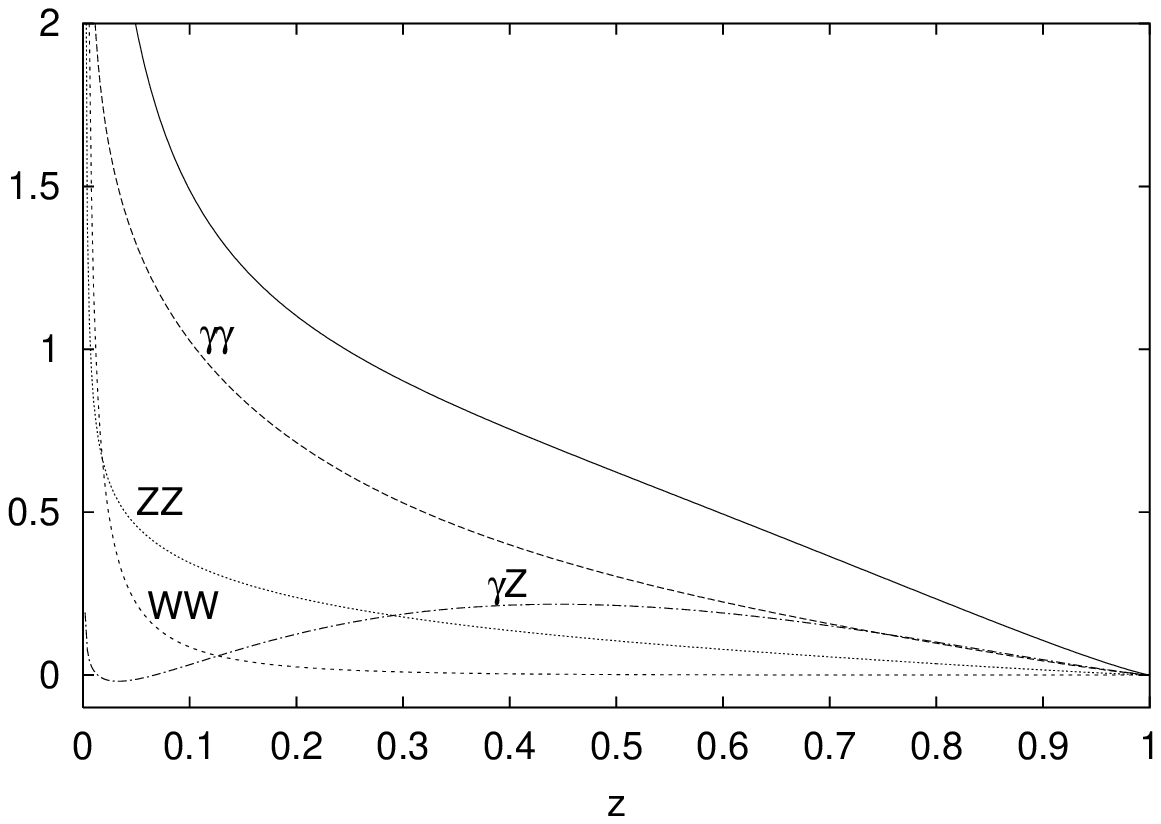,width=\columnwidth}%
\unitlength=1mm
\put(-25,45){$z u^{\rm as}(z)$}}
\centerline{\epsfig{file=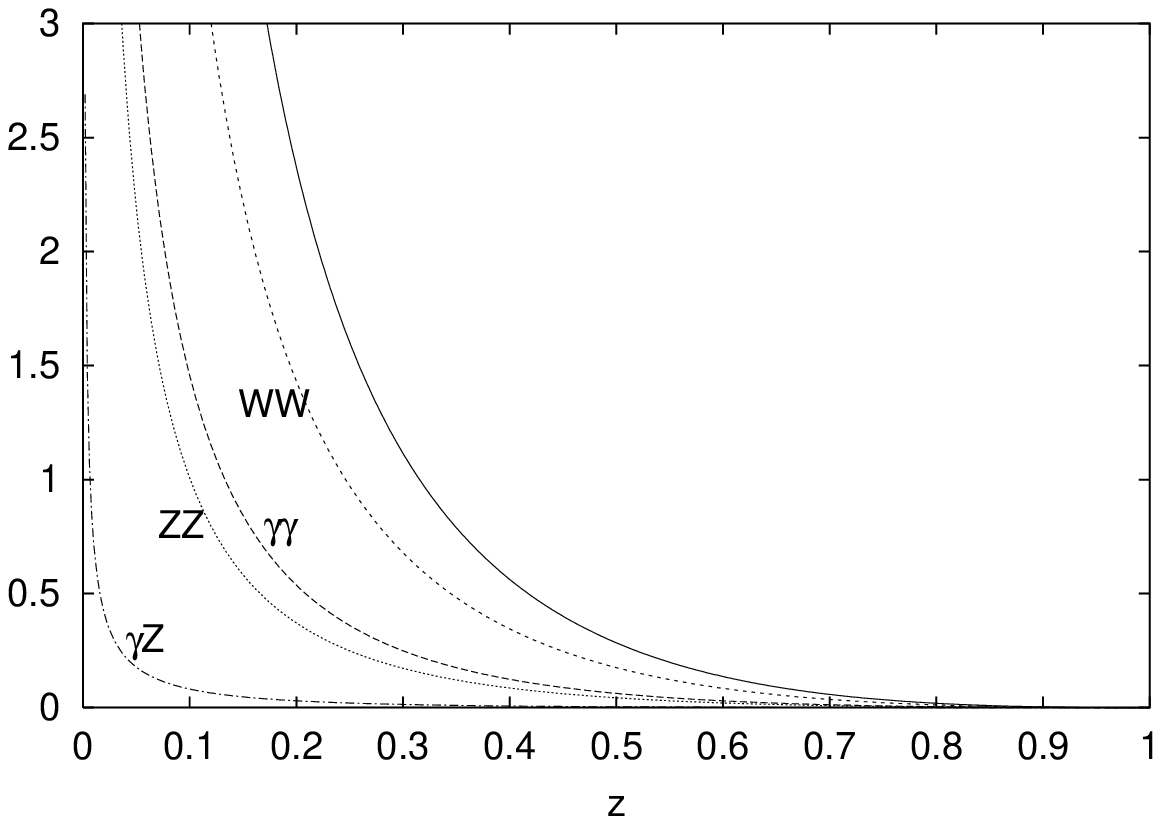,width=\columnwidth}%
\unitlength=1mm
\put(-25,45){$z G^{\rm as}(z)$}}
%\vspace{2mm}
\caption{Asymptotic quark and gluon distributions  --- solid line. 
The other lines show different contributions as labeled}
\end{figure}

The 
`inclusive' (no tagging)
case when $a=1$
was considered in Ref. \cite{WS22,wojtus}. 
After quite non-trivial procedure 
the evolution equations 
look formally identical to Eqs.(\ref{master}) with $t_1$ replaced by $t$.
In particular the  splitting function $P^e_q$ is now proportional to $t$.
The solutions again have the form (\ref{asdef}) with $t_1$ replaced by $t$
(the parton densities grow thus as $t^2$),
but the resulting 
$z$-dependence is different
from the previous case.
The standard convolution of equivalent photons with the photon structure 
(as in case $a=0$) is no more valid.

The comparison of the two cases is shown for the electron 
structure function $F_2^e(z,Q^2,P^2)$ in Fig. 4. 
Two upper curves are the asymptotic result 
with all electroweak bosons taken into account.
We also show the photon contribution alone to get a feeling
what is the effect at presently accessible momenta. 
One sees an extra suppression in the `inclusive' case (when $t_1 \sim t$).

Let us add a few final remarks. First concerns the study of the
virtual photon structure \cite{virt}. The analysis can be reformulated 
in terms of the $P^2$ dependence of the electron structure function. 
Studying a real, convention independent object is first advantage. 
Another one is the fact that
at very high virtualities the $Z$ admixture and the $\gamma$-$Z$
interference are properly taken into account.

%%%%%%%  fig.4 **********************************
\begin{figure}
\centerline{\epsfig{file=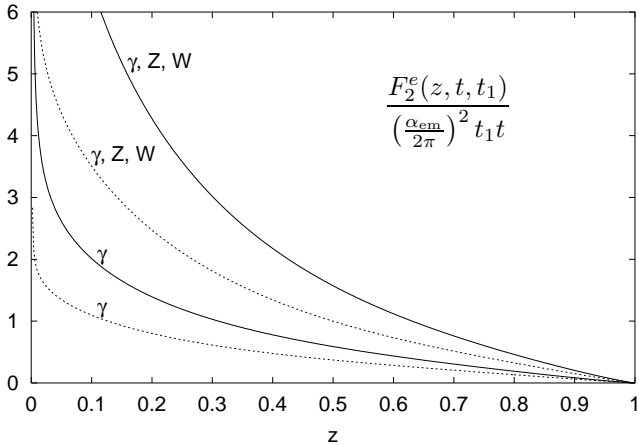,width=\columnwidth}%
\unitlength=1mm
\put(-35,47){$\displaystyle
{F_2^e(z,t,t_1) \over \left({
%\displaystyle
\alpha_{\rm em} \over 2\pi}\right)^2%}
%\log {P^2 \over \Lambda_{\rm QCD}^2} \log {Q^2 \over\Lambda_{\rm QCD}^2}
t_1t
}
$}}
%\vspace{2mm}
\caption{$F_2^e(z,Q^2,P^2) / ({\alpha_{\rm em} \over 2\pi})^2
%\log {P^2 \over \Lambda^2} \log {Q^2 \over\Lambda^2}
t_1t
$ for $P^2 \ll Q^2$ --- solid line,
and for $P^2 \approx Q^2$ --- broken line. 
The contributions from $\gamma, Z, W$ and from $\gamma$ alone are labeled}
\end{figure}

Second,   
the calculation discussed above has been 
done in leading logarithmic approximation.
Corrections to this picture are well defined. In particular one does not
require factorization of the electron induced process into the $e$-$\gamma$
(Weizs\" acker-Williams) and $\gamma$-QCD parts. The process initializing
the $q\bar q$ cascade can be 
calculated exactly (in given order of $\alpha_{\rm em}$).

Third, the extraction of the electron structure function consists in the 
reanalysis of the collected $e^+e^-$ data which served for the photon.
A phenomenological parametrisation describing the data, with the predicted
$Q^2$ and $P^2$ dependence, could
be then constructed (including the VDM and direct contributions, see also \cite{DG}). 
This parametrisation would serve then as the basis for a  
Monte-Carlo generator
simulating quark production in lepton induced processes. This single 
generator 
replaces in fact two others, presently used\cite{M-C}.

As a side-remark we comment on the QED structure function of the photon.
It is extracted from the process $e^+e^- \rightarrow e^+e^-\mu^+\mu^-$
by dividing out the (approximate) equivalent photon distribution.
The use of the QED electron structure function avoids this problem.
The exactly known (in given order of $\alpha_{\rm em}$) electron 
structure function can be compared directly with the electron data.

Finally, the photon structure function has been recently measured 
\cite{dijets} in two jet production at HERA. Again the extraction of the $x$
variable is difficult. In addition to jets, one has to measure 
essentially  the whole hadronic system in order to obtain the photon 
energy. The data, when presented in terms of the electron 
structure function, require only measurement of the two jets. As in 
the $e^+e^-$ case the resulting data should be more precise.

To summarize, we propose to look at the electron as surrounded by a
QCD cloud of quarks and gluons (in order $\alpha_{\rm em}^2$), very much
like it is surrounded by a QED cloud of equivalent photons (in order $\alpha_{\rm em}$).
We argue that the use of the electron structure function
in electron induced processes has important advantages over the photon
one. Experimentally it 
leads to more precise, convention independent data. Theoretically
it allows for more careful treatment of all variables. It  also takes 
into account all electroweak gauge boson contributions, including their interference, which will
be important in the next generation of $e^+e^-$ colliders.

The authors would like to thank Maria Krawczyk, Aharon Levy and 
Jacek Turnau for discussions. The hospitality of the DESY Theory 
Group is also acknowledged.

\end{document}